\documentclass[aps,prl,preprint,tightenlines,superscriptaddress,showpacs,byrevtex]{revtex4}

\usepackage{graphicx}
\usepackage{dcolumn}  
\usepackage{epsfig}
\usepackage{amsmath}
\usepackage{refmerge}

\def\um{\ifmmode {\mathrm{\ \mu m}}\else
                  \textrm{$\mu$m}\fi}%
\def\GeV{\ifmmode {\mathrm{\ Ge\kern -0.1em V}}\else
                   \textrm{Ge\kern -0.1em V}\fi}%
\def\MeV{\ifmmode {\mathrm{\ Me\kern -0.1em V}}\else
                   \textrm{Me\kern -0.1em V}\fi}%
\def\keV{\ifmmode {\mathrm{\ ke\kern -0.1em V}}\else
                   \textrm{ke\kern -0.1em V}\fi}%
\def\eV{\ifmmode  {\mathrm{\ e\kern -0.1em V}}\else
                   \textrm{e\kern -0.1em V}\fi}%

\def\GeVc{\ifmmode  {\mathrm{\ Ge\kern -0.1em V/c}}\else
                    \textrm{Ge\kern -0.1em V/c}\fi}%
\def\MeVc{\ifmmode  {\mathrm{\ Me\kern -0.1em V/c}}\else
                    \textrm{Me\kern -0.1em V/c}\fi}%

\def\GeVcsq{\ifmmode  {\mathrm{\ Ge\kern -0.1em V/c^2}}\else
                      \textrm{Ge\kern -0.1em V/c$^2$}\fi}%
\def\MeVcsq{\ifmmode  {\mathrm{\ Me\kern -0.1em V/c^2}}\else
                      \textrm{Me\kern -0.1em V/c$^2$}\fi}%

\def\GeVsqcsq{\ifmmode  {\mathrm{\ Ge\kern -0.1em V^2/c^2}}\else
                      \textrm{Ge\kern -0.1em V$^2$/c$^2$}\fi}%
\def\MeVsqcsq{\ifmmode  {\mathrm{\ Me\kern -0.1em V^2/c^2}}\else
                      \textrm{Me\kern -0.1em V$^2$/c$^2$}\fi}%

\def\BW{{\mathrm{B\kern -0.2em W \kern -0.2em}}}
\def\DK{{\mathrm{D\kern -0.2em K \kern -0.1em}}}
\def\NN{{\mathrm{N\kern -0.2em N}}}
\def\BR{{\ B\kern -0.2em r \kern -0.1em}}
\def\Br{{\mathcal{B}}}

\def\ndf{{\mathrm{ndf}}}
\def\MC{{\mathrm{MC}}}

\newcommand{\ppp}{\pi^+\pi^-\pi^0 }
\newcommand{\Ggg}{\Gamma_{\gamma\gamma} }
\newcommand{\Wgg}{W_{\gamma\gamma} }

\begin{document}

\preprint{\vbox{ \hbox{ }
                 \hbox{BELLE-CONF-0662 }
                 \hbox{08 October 2006 } 
}}

\title{ \quad\\[3.0cm] 
    Study of tensor states 
    in the reaction $\gamma\gamma \rightarrow \pi^+\pi^-\pi^0$
    \quad\\[0.5cm]
}
\affiliation{Budker Institute of Nuclear Physics, Novosibirsk}
\affiliation{Chiba University, Chiba}
\affiliation{Chonnam National University, Kwangju}
\affiliation{University of Cincinnati, Cincinnati, Ohio 45221}
\affiliation{University of Frankfurt, Frankfurt}
\affiliation{The Graduate University for Advanced Studies, Hayama} 
\affiliation{Gyeongsang National University, Chinju}
\affiliation{University of Hawaii, Honolulu, Hawaii 96822}
\affiliation{High Energy Accelerator Research Organization (KEK), Tsukuba}
\affiliation{Hiroshima Institute of Technology, Hiroshima}
\affiliation{University of Illinois at Urbana-Champaign, Urbana, Illinois 61801}
\affiliation{Institute of High Energy Physics, Chinese Academy of Sciences, Beijing}
\affiliation{Institute of High Energy Physics, Vienna}
\affiliation{Institute of High Energy Physics, Protvino}
\affiliation{Institute for Theoretical and Experimental Physics, Moscow}
\affiliation{J. Stefan Institute, Ljubljana}
\affiliation{Kanagawa University, Yokohama}
\affiliation{Korea University, Seoul}
\affiliation{Kyoto University, Kyoto}
\affiliation{Kyungpook National University, Taegu}
\affiliation{Swiss Federal Institute of Technology of Lausanne, EPFL, Lausanne}
\affiliation{University of Ljubljana, Ljubljana}
\affiliation{University of Maribor, Maribor}
\affiliation{University of Melbourne, Victoria}
\affiliation{Nagoya University, Nagoya}
\affiliation{Nara Women's University, Nara}
\affiliation{National Central University, Chung-li}
\affiliation{National United University, Miao Li}
\affiliation{Department of Physics, National Taiwan University, Taipei}
\affiliation{H. Niewodniczanski Institute of Nuclear Physics, Krakow}
\affiliation{Nippon Dental University, Niigata}
\affiliation{Niigata University, Niigata}
\affiliation{University of Nova Gorica, Nova Gorica}
\affiliation{Osaka City University, Osaka}
\affiliation{Osaka University, Osaka}
\affiliation{Panjab University, Chandigarh}
\affiliation{Peking University, Beijing}
\affiliation{University of Pittsburgh, Pittsburgh, Pennsylvania 15260}
\affiliation{Princeton University, Princeton, New Jersey 08544}
\affiliation{RIKEN BNL Research Center, Upton, New York 11973}
\affiliation{Saga University, Saga}
\affiliation{University of Science and Technology of China, Hefei}
\affiliation{Seoul National University, Seoul}
\affiliation{Shinshu University, Nagano}
\affiliation{Sungkyunkwan University, Suwon}
\affiliation{University of Sydney, Sydney NSW}
\affiliation{Tata Institute of Fundamental Research, Bombay}
\affiliation{Toho University, Funabashi}
\affiliation{Tohoku Gakuin University, Tagajo}
\affiliation{Tohoku University, Sendai}
\affiliation{Department of Physics, University of Tokyo, Tokyo}
\affiliation{Tokyo Institute of Technology, Tokyo}
\affiliation{Tokyo Metropolitan University, Tokyo}
\affiliation{Tokyo University of Agriculture and Technology, Tokyo}
\affiliation{Toyama National College of Maritime Technology, Toyama}
\affiliation{University of Tsukuba, Tsukuba}
\affiliation{Virginia Polytechnic Institute and State University, Blacksburg, Virginia 24061}
\affiliation{Yonsei University, Seoul}
  \author{K.~Abe}\affiliation{High Energy Accelerator Research Organization (KEK), Tsukuba} 
  \author{K.~Abe}\affiliation{Tohoku Gakuin University, Tagajo} 
  \author{I.~Adachi}\affiliation{High Energy Accelerator Research Organization (KEK), Tsukuba} 
  \author{H.~Aihara}\affiliation{Department of Physics, University of Tokyo, Tokyo} 
  \author{D.~Anipko}\affiliation{Budker Institute of Nuclear Physics, Novosibirsk} 
  \author{K.~Aoki}\affiliation{Nagoya University, Nagoya} 
  \author{T.~Arakawa}\affiliation{Niigata University, Niigata} 
  \author{K.~Arinstein}\affiliation{Budker Institute of Nuclear Physics, Novosibirsk} 
  \author{Y.~Asano}\affiliation{University of Tsukuba, Tsukuba} 
  \author{T.~Aso}\affiliation{Toyama National College of Maritime Technology, Toyama} 
% ...
\collaboration{The Belle Collaboration}

\collaboration{Belle Collaboration }
\noaffiliation

\begin{abstract}
\noindent
   A high statistics study of the reaction  
   $\gamma\gamma\to \pi^+\pi^-\pi^0$ has been performed
   with the Belle detector using a data sample of 26 fb$^{-1}$ 
   collected at $\sqrt{s}=10.58\GeVcsq$.
   A spin-parity analysis shows dominance of the $J^P=2^+$ 
   helicity 2 wave for three-pion invariant masses 
   from 1 to 3 \GeVcsq.
   The invariant mass distribution exhibits $a_2(1320)$, 
   $a_2(1700)$ and higher mass enhancements.

\end{abstract}


\maketitle        
\tighten
\normalsize


\section {1. Introduction}  

Three-pion final states of two-photon interactions are restricted to 
quantum numbers suitable for study of resonance formation.
The $\rho\pi$ channel is known to be dominated by the formation of 
$a_2(1320)$ in the helicity 2 state [1-10].
The $a_2(1320)$ is a ground state of isospin 1 $^3P_2$ $q\bar{q}$ 
mesons.  Observations of higher mass resonances have been reported
[7-10].  
Study of higher mass states is important for the assignment of nonet 
members and for the understanding of confinement in the quark model 
[11-15].  
The L3 collaboration reported the $a_2(1700)$ in 
$\rho\pi$ and $f_2\pi$ decay modes mainly in helicity 2 
states \cite{L3}.  The Belle observation of a resonance in 
$\gamma\gamma\to K^+K^-$ \cite{BelleKK} at 1737 \MeVcsq{} is listed
in the Particle Data Group (PDG) \cite{PDG} and attributed to the 
$a_2(1700)$.
Experiments on $\pi p$ collisions [18-20] 
and $\gamma p$ photo-production \cite{Aston,Condo} have also 
observed a three-pion resonance near $1.8 \GeVcsq$.
The Crystal Barrel collaboration reported the $a_2(1700)$ in the 
$\pi^0\eta$ mode in $\bar{p} p$ collisions 
\cite{CBAR02,CBAR99}.
The $a_2(2100)$ and $a_2(2280)$ in the $f_2 \pi$ mode 
were reported in \cite{CBAR_AVA}.

We present a spin-parity analysis of three-pion events using data 
collected with the Belle detector at KEKB \cite{KEKB}.
Resonance formation is investigated in the $\rho\pi$ and 
$f_2\pi$ modes, including interference between them.
The data sample was taken with similar trigger conditions during 
2000 and 2001 at a center-of-mass energy $\sqrt{s}=10.58 \GeV$.
The corresponding total integrated luminosity is 26.0 fb$^{-1}$.
We first describe the theoretical formulae for spin-parity 
analysis of three-pion events.
The spin dependence is investigated using the distribution of 
the vector normal to the three-pion decay plane.
Upper limits on two-photon radiative widths are determined for 
$\pi_0(1300)$, $\pi_2(1670)$ and $a_4(2040)$.  
A neural network method is applied to enhance tensor state selection.
The helicity states of tensor resonances are investigated 
using $\cos\theta$ distributions of final state pions.
Background contamination is evaluated by comparing $p_t^2(3\pi)$ 
distributions for data and Monte Carlo generated two-photon 
interactions.
The three-pion and di-pion mass spectra are investigated.
The two-photon radiative width of the $a_2(1320)$ is measured and
compared to the PDG world average.  The mass region above $1.5 \GeVcsq$ 
is examined for the $a_2(1700)$ and new states.
The two-photon radiative widths are determined.
Interference between tensor states and $\rho\pi$ and $f_2\pi$ 
decay modes is investigated and the coupling amplitudes and 
phase angles are also determined.

\section {2. Theoretical formulae and Monte Carlo}
\label{sec:mc}

The three-pion final state in two-photon interactions is expressed
as the composition of production of di-pion isobars, $I \pi$, 
and the decay of $I \rightarrow \pi\pi$.
Possible di-pion isobars are $f_0$, $\rho^\pm$ and $f_2$.  
The assignment of spin-parity ($J^P$) and orbital angular momentum (L)
are listed in Table~\ref{tab:spin}.
Gauge invariance and Bose symmetry forbid $J^P =1^\pm$ and $3^-$.
Only the helicity $\lambda=0$ state is allowed for $J^P=0^-$ and $2^-$,
and $\lambda=2$ for $J^P=3^+$ \cite{Yang}.
The helicity 0 fraction of a tensor state is predicted to be zero by 
the non-relativistic quark model \cite{Li,Grassberger}.
The static quark model also predicts a small value (1/7) \cite{Poppe}.

\begin{table}[t!]
  \begin{center} \begin{tabular}{lccccc}
  \hline
  \hline
   $J^P$      &  $0^-$  &  $2^-$  &  $2^+$  &  $3^+$  &  $4^+$  \\
  \hline
   $\lambda$  &    0    &   0     &   0,2   &   2     &   0,2  \\
  \hline
   $L$ ($f_0\pi^0$)
              &    0    &   2     &    -    &   3     &    -   \\
   $L$ ($\rho^\pm\pi^\mp$)
              &    1    &   1     &    2    &   2     &    4   \\
   $L$ ($f_2\pi^0$)
              &    2    &   0     &    1    &   1     &    3   \\
  \hline
  \hline
  \end{tabular}
  \caption{ Spin-parity of the three-pion final state
     in two-photon interactions.
     Listed are the possible di-pion isobars ($f_0$, $\rho$ or $f_2$),
     helicity ($\lambda$) states allowed,
     and the lowest orbital angular momentum ($L$) between the 
     di-pion isobar and the third pion.
  \label{tab:spin}}
  \end{center}
  \vspace{-.5cm}
\end{table}

The cross section for resonance formation in 
$\gamma\gamma \rightarrow I\pi$ is given by
\begin{eqnarray} 
 \lefteqn { d \sigma_{\gamma\gamma}  =
  2\pi (2 J +1)  
  \Gamma_{\gamma\gamma} \sum_{J_z} R_{J_z}
  \left| 
  \sqrt{\frac{m_0}{s}} \BW_0  \! \sum_I \! D_0^{J_z}(I) \right. }
  \nonumber  \\
  && \left. 
        + \alpha_1 e^{i\phi_1}  
          \sqrt{\frac{m_1}{s}} \BW_1  \! \sum_I \! D_1^{J_z}(I)
        + \alpha_2 e^{i\phi_2} 
          \sqrt{\frac{m_2}{s}} \BW_2 \! \sum_I \! D_2^{J_z}(I)
        + \! \cdot\cdot \right |^2 d {\rm Lips}(3\pi), 
\label{eq:xsec}
\end{eqnarray} 
where $\Ggg$ is the two-photon radiative width of the
ground state resonance of spin $J$ and
$R_{J_z}$ is the probability of helicity $J_z$. 
$\BW_i=1/( s - m_i^2 +i m_i \Gamma_i)$ is the Breit-Wigner 
term for two-photon invariant mass $(\Wgg=\sqrt{s})$
for a resonance of mass $m_i$ and width $\Gamma_i$.
The decay modes are denoted by $D_i^{J_z}(I)$.
Interference with the ground state is given 
by the coupling amplitude $\alpha_i$ and phase angle $\phi_i$,
where the radiative width is $\alpha_i^2 \Gamma_{\gamma\gamma}$.
The decay amplitude is given by
\begin{equation} 
   \sum_I D_i^{J_z}(I)=
        \BW(\rho^+) T^{J_z}(\rho^+,\pi)
      + \BW(\rho^-) T^{J_z}(\rho^-,\pi) 
      + \xi_i e^{i\psi_i }\BW(f_2) T^{J_z}(f_2,\pi) 
\label{eq:di}
\end{equation}
for the Breit-Wigner terms of $\rho$ and $f_2$ isobars
with interference expressed by the amplitude $\xi_i$ 
and phase angle $\psi_i$.
The spin dependence is given by 
\begin{equation}     
  T^{J_z} (I, \pi)   
 =  32\pi^2 \left( m_R \Gamma_R \, m_{2\pi} \Gamma_{2\pi}
 \frac { \sqrt { s \, s_{2\pi} } }
       { p_{2\pi} \, p_\pi} \right)^{1/2}  
     \sum_m
         C^{J,J_z}_{L,J_z-m,l,m}
         Y^{J_z-m}_L (\theta_{2\pi},\phi_{2\pi} )
         Y^m_l ( \theta_\pi, \phi_\pi),
\end{equation}
where the subscript $2\pi$ denotes
parameters of the the di-pion ($\rho^\pm$ or $f_2$) isobar 
with spin $l$ and third component $m$.
The subscript $\pi$ denotes a pion from the di-pion decay.
The spherical harmonics are multiplied by the Clebsch-Gordan 
coefficients $C^{J,\lambda}_{L,\lambda-m,l,m}$. They describe
the angular distribution of the di-pions (with invariant mass 
$m_{2\pi}=\sqrt{s_{2\pi}}$, momentum $p_{2\pi}$, polar angle
$\theta_{2\pi}$, and azimuthal angle $\phi_{2\pi}$) in the
resonance rest frame and the pions from the di-pion decay (with
$p_\pi,\theta_\pi$, $\phi_\pi$) in the di-pion rest frame.
The $\theta$ and $\phi$ angles in both frames are determined
with respect to the incident
$\gamma\gamma$ direction which is approximated by the $e^+e^-$ beam
direction in the center-of-mass frame.

The Monte Carlo is prepared for two-photon interactions in the 
asymmetric beam configuration (8 \GeV{} $e^-$ on 3.5 \GeV{} $e^+$) 
at Belle.
The incident photon flux is calculated with the two-photon luminosity 
function in \cite{Budnev}. 
The colliding photons are generated with the $\rho$-pole form factor
of the vector meson dominance model (VDM).
The photon flux has been compared to the QED calculation of the 
{\sf DIAG36} program \cite{DIAG36}.
The systematic uncertainty on the photon flux simulation is about
 2\%.

The detector response is simulated with GEANT3 \cite{GEANT}.
The event trigger is simulated using trigger inputs from 
the Belle tracking system and electromagenetic calorimeter.

\section {3. Event selection } 

Events are pre-selected for two-photon interactions
with the energy sum of final state particles below 5 \GeV{}
and the scattered electrons not observed.
Charged pions are detected by the central drift chamber (CDC) and 
the silicon vertex detector (SVD) in a 1.5 T magnetic field.
The coverage in polar angle extends from $17^\circ$ to $150^\circ$ 
in the laboratory frame. 
Outside the CDC are the Aerogel \v{C}erenkov Counter system (ACC) 
and the Time Of Flight (TOF) system.
The photons from $\pi^0$ decays are detected by the electromagnetic 
calorimeter (ECL) comprised of CsI(Tl) crystals covering 
the same angular region as the CDC. A detailed description of the 
detectors and their performance is given in reference \cite{belle}.


The three-pion event trigger is composed of a CDC 
two-track trigger, an ECL energy trigger, and combinations with 
looser constraints on the track opening angles and energy thresholds.
In event reconstruction, charged pion tracks are required to have
transverse momentum larger than 300 \MeVc.
The purity of the charged pions is enhanced by  rejecting
electrons using a likelihood function based on  
the ratio of ECL energy to track momentum and $dE/dx$ of the track
in the CDC
\cite{ElectronID}. 
Charged kaons are rejected by cuts on a probability function 
based on ACC, TOF and CDC measurements.
ECL and beam background are suppressed by
requiring a maximum distance to the interaction point (IP) of 2.5~mm 
in the $r$-$\phi$ plane and 50~mm along $z$ (the beam direction). 
A photon is identified as an isolated ECL cluster with no matching 
track.  Partially contained photons in the ECL and beam background are
eliminated by requiring an energy threshold $E_\gamma$ above 100 \MeV.

Three-pion candidates are selected by requiring two
oppositely charged pions and a photon pair with invariant
mass within three sigma of the $\pi^0$ mass.
Background events are suppressed by a tight cut on the 
three-pion transverse momentum and
a high energy threshold for photons from $\pi^0$ decay.
Additional selection criteria are chosen for the event topology.
Distributions of charged pions and photons are simulated for 
the partial waves and decay modes to be investigated.
In the three-pion rest frame, the photons from $\pi^0$ decay are 
expected to have a large opening angle ($\Omega(\gamma\gamma)$)
and to make large angles with the charged pions ($\Omega(\pi\gamma)$).
Background is expected from inclusive events or low 
multiplicity events contaminated with random photons.
These events populate the low opening angle region
and can be eliminated 
by requiring $\min(\Omega(\pi\gamma) + \Omega(\gamma\gamma))> 1$ radian.
Background events with a di-pion in the low mass region outside 
the Dalitz contour are excluded by 
$m(\pi^+\pi^0) + m(\pi^-\pi^0) >1 \GeVcsq$.
Contamination by $\gamma\gamma \rightarrow
\pi^+\pi^-$ events in coincidence with a fake $\pi^0$ is
suppressed by requiring $p_t^2(\pi^+\pi^-)>0.01 \GeVsqcsq$,
as the transverse momentum of the $\pi^+\pi^-$ pair is 
balanced in direct production.

\begin{figure}[b!]
  \hspace{.5cm}
  \epsfig{file=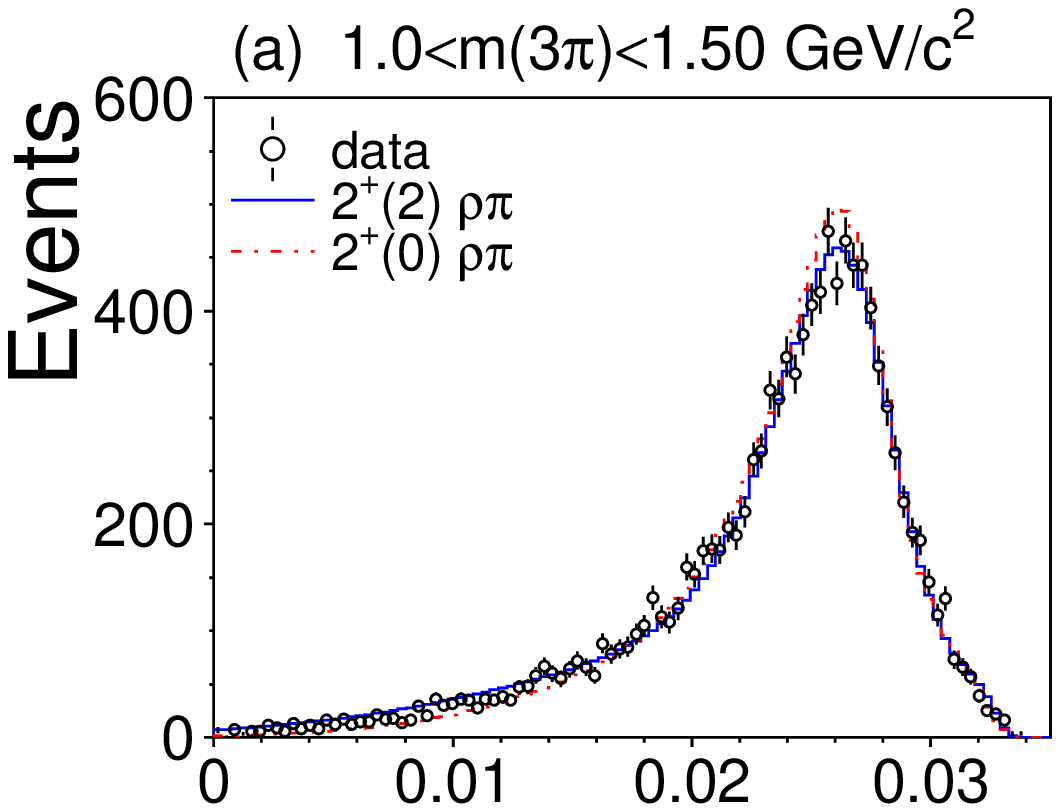,width=.26\linewidth}
  \hspace{-1.cm}
  \epsfig{file=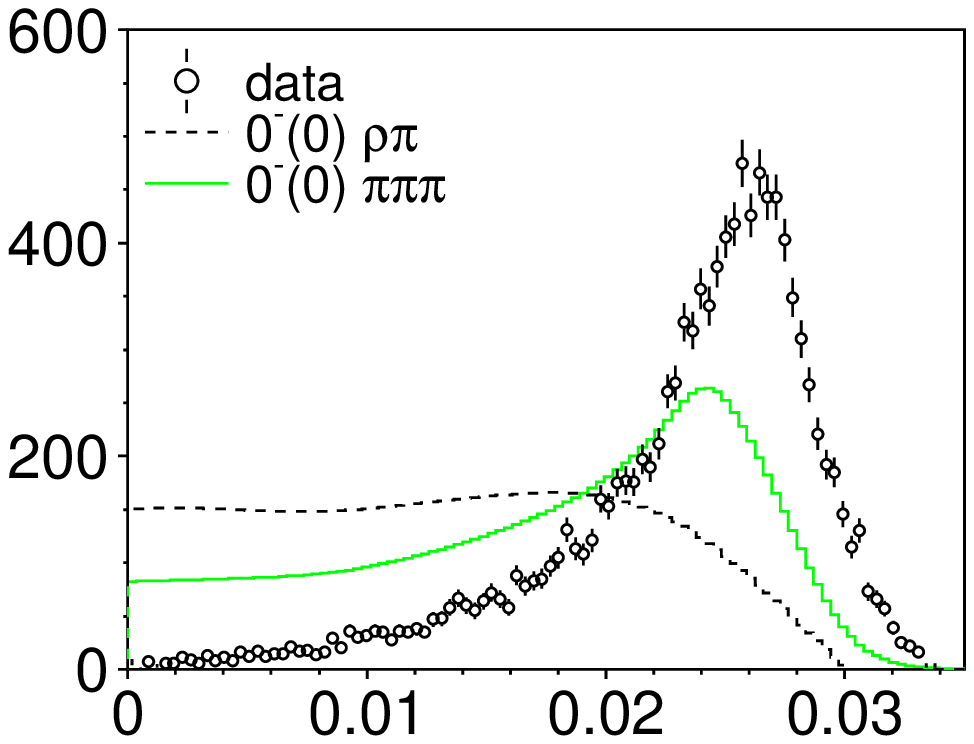,width=.26\linewidth}

  \vspace{-.2cm}

  \hspace{.5cm}
  \epsfig{file=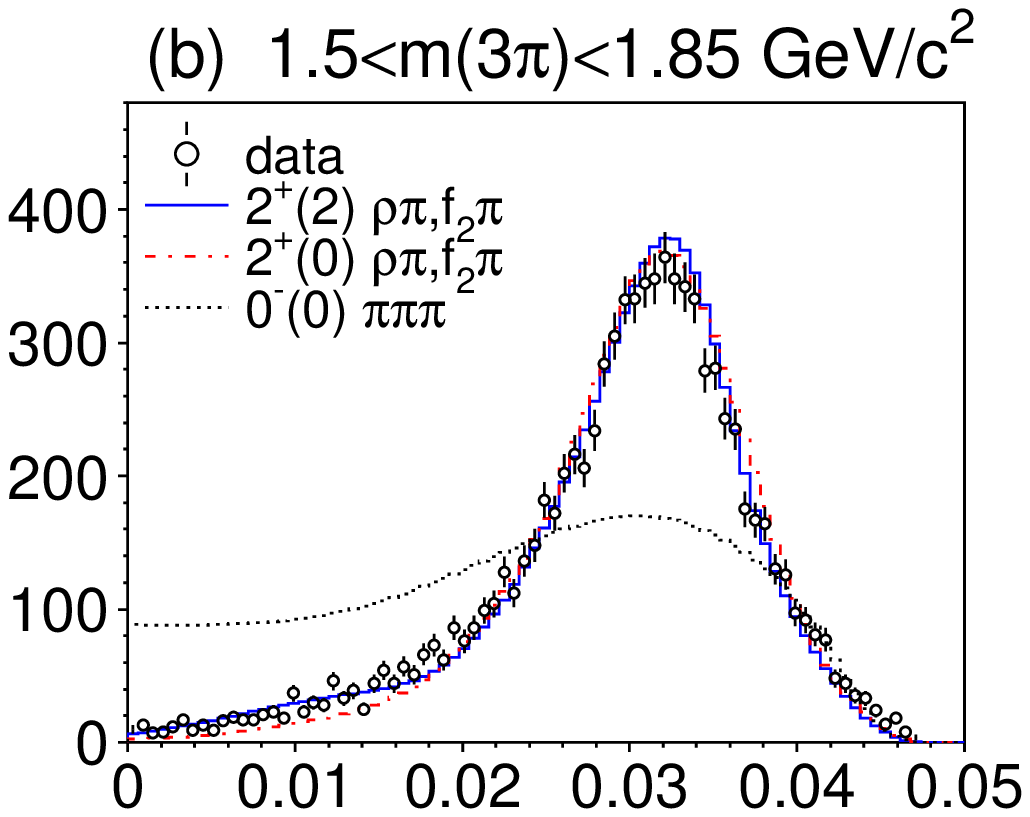,width=.26\linewidth}
  \hspace{-1.cm}
  \epsfig{file=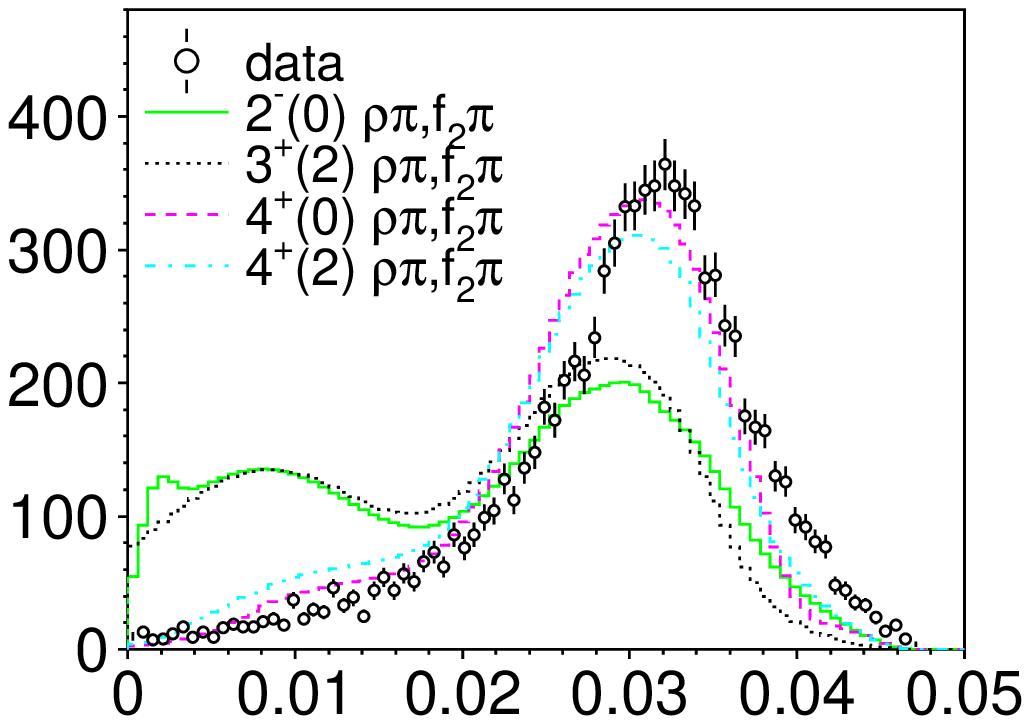,width=.26\linewidth}

  \vspace{-.2cm}

  \hspace{.5cm}
  \epsfig{file=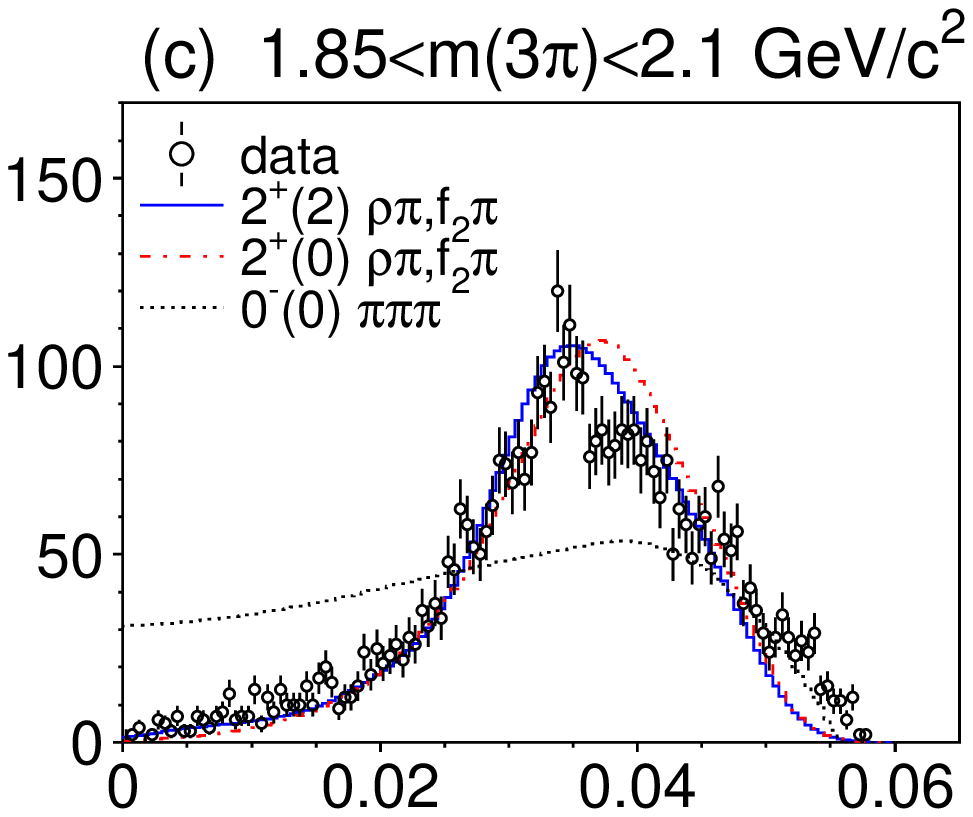,width=.26\linewidth}
  \hspace{-1.cm}
  \epsfig{file=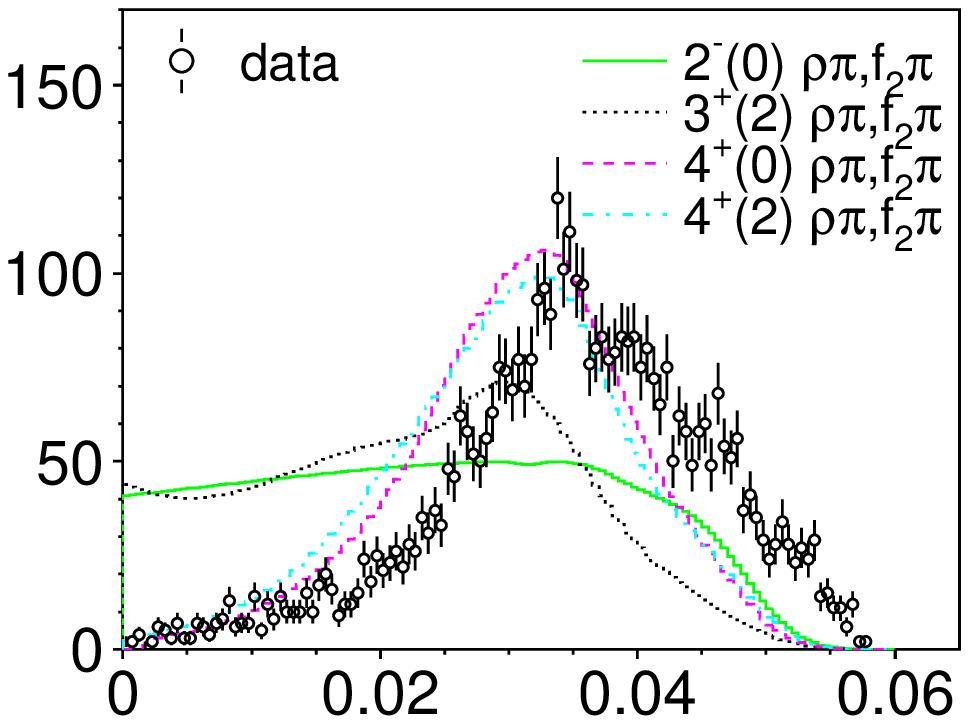,width=.26\linewidth}

  \vspace{-.2cm}

  \hspace{.5cm}
  \epsfig{file=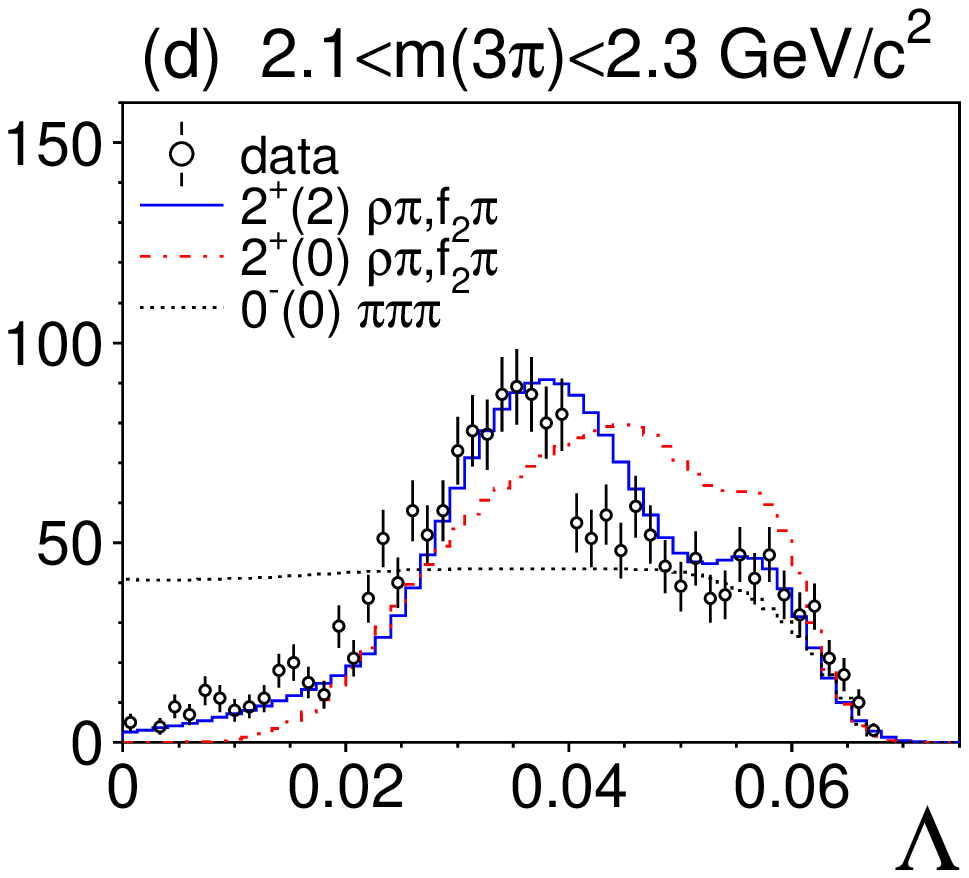,width=.26\linewidth}
  \hspace{-1.cm}
  \epsfig{file=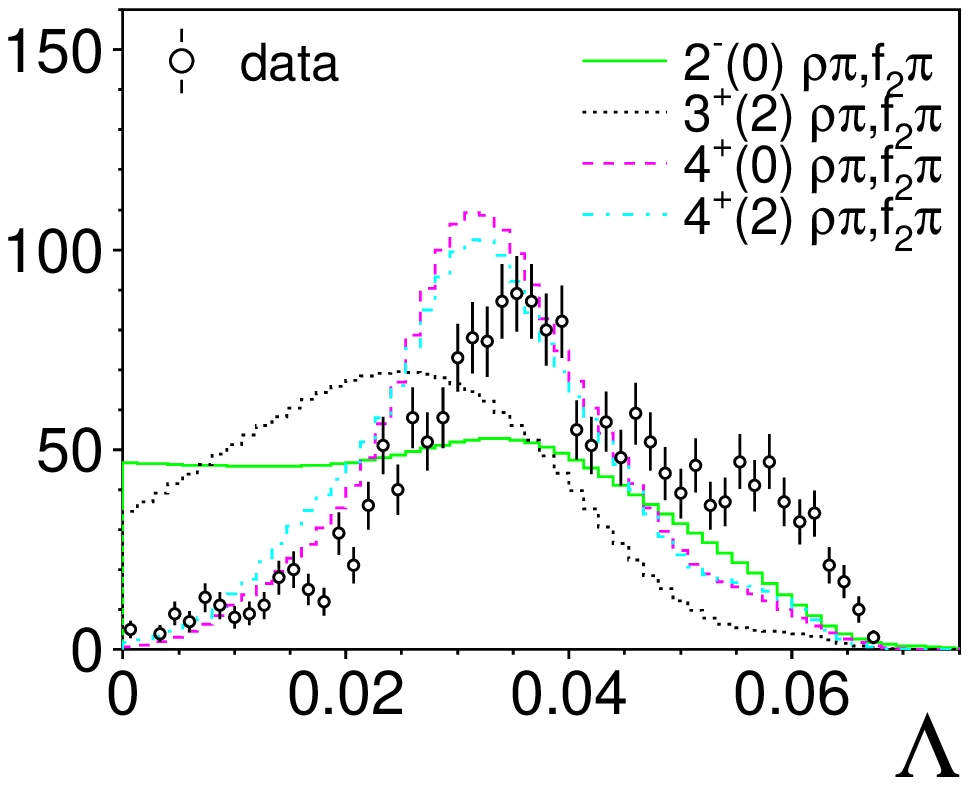,width=.26\linewidth}

  \vspace{-.2cm}
  \caption{$\Lambda$ distributions in comparison with
      Monte Carlo of various spin-parity hypotheses. 
      Monte Carlo distributions in each mass interval
      are simulated for resonances of mass and width 
      listed in Table~\ref{tab:lambda}.
  \label{fig:nnvar_npm} }
  \vspace{-.5cm}
\end{figure}

\begin{table}[t!]
  \begin{center} \begin{tabular}{cc|cccc}
  \hline
   \multicolumn{2}{c|}{ $m(3\pi)$ range (\GeVcsq) }
                        & 1.0-1.5 & 1.5-1.85 & 1.85-2.1 & 2.1-2.3  \\
  \hline
   \multicolumn{2}{r|}{ MC resonance mass (\MeVcsq) }
                        & 1318    & 1750     & 1950     & 2140   \\
   \multicolumn{2}{r|}{             width (\MeVcsq) }
                        &  105    &  250     &  250     &  250   \\
  \hline
   \multicolumn{2}{c|}{ Decay modes, $J^P$(helicity) }&
   \multicolumn{4}{c}{ $\chi^2$/ndf} \\
   $\ppp$              &  phase space  & 3540/91  &  2050/76  &  716/101   & 338/47 \\
   $\rho^\pm\pi^\mp$   &  $0^-(0)$     & 7790/83  &  4510/70  & 1670/82   & 733/31 \\
   $\rho^\pm\pi^\mp$   &  $2^+(0)$     &  101/83  &  -  &  -    &  -  \\
   $\rho^\pm\pi^\mp$   &  $2^+(2)$     &   73/93  &  -  &  -    &  -  \\
   $\rho^\pm\pi^\mp$   &  $2^-(0)$     &  -   & 1760/76  &   565/99  &  220/45 \\
   $f_2\pi^0$          &  $0^-(0)$     &  -   & 4890/77  &  1100/102 &  760/48 \\
   $f_2\pi^0$          &  $2^-(0)$     &  -   & 3540/76  &   818/105 &  323/48 \\
   $\rho\pi + f_2\pi^0$ &  $2^+(0)$    & -    &  134/70  &   125/83  &  127/38 \\
   $\rho\pi + f_2\pi^0$ &  $2^+(2)$    & -    &   71/76  &   120/89  &   62/45 \\
   $\rho\pi + f_2\pi^0$ &  $2^-(0)$    & -    & 2770/75  &  1170/95  &  427/46 \\
   $\rho\pi + f_2\pi^0$ &  $3^+(2)$    & -    & 3090/72  &  1520/90  &  596/40 \\
   $\rho\pi + f_2\pi^0$ &  $4^+(0)$    & -    &  425/71  &   450/89  &  212/42 \\
   $\rho\pi + f_2\pi^0$ &  $4^+(2)$    & -    &  504/74  &   542/92  &  210/43 \\
  \hline
  \end{tabular}
  \caption{$\chi^2$/ndf of goodness-of-fit tests to the 
     $\Lambda$ distributions of spin-parity and helicity states.
     Monte Carlo simulations were performed 
     with  the resonance masses and widths listed.
  \label{tab:lambda} }
  \end{center}
  \vspace{-.5cm}
\end{table}

\section {4. Spin-parity analysis }

The spin-parity of a three-pion event is characterized by 
the spherical harmonics in Eq.~\ref{eq:di}.
The dependence is explicitly seen in the angular distribution 
of the final state pions and of the vector normal 
to the three-pion decay plane.
The spin-parity of the data sample is identified using
the $\Lambda$ parameter, which is the squared magnitude of the 
normal vector, scaled by the maximum available kinetic energy 
\cite{Pluto84,L3},
\begin{equation}
  \Lambda=\left| \vec{N}/Q \right|^2
\end{equation}
where $\vec{N}=\vec{p}_{\pi^+} \times\vec{p}_{\pi^-}$ is 
evaluated in the three-pion rest frame and $Q$ is the rest energy 
difference between the resonance and the three pions.
The normal vector $\vec{N}$ covers a wider $\cos\theta$ range than 
the detector acceptance, providing better discrimination than 
the the $\cos\theta$ values of individual pions.

Goodness-of-fit tests to the $\Lambda$ 
distributions in Fig.~\ref{fig:nnvar_npm} were performed in four 
invariant mass intervals.  The event sample is selected with 
$p_t^2(3\pi) < 0.001 \GeVsqcsq$ and $E_\gamma>150 \MeV{}$.
The mass intervals were chosen to select the $a_2(1320)$, 
the $a_2(1700)$ and possible higher mass states.
Monte Carlo distributions were generated for resonances
of each partial wave hypothesis investigated.
The generated resonance masses and widths correspond to the 
enhancements observed in data (to be discussed in Section 7).
Interference between the $\rho\pi$ and $f_2\pi^0$ channels is 
included using $\xi=0.91$ and $\psi=150^\circ$.

The $a_2(1320)$ is dominant in the mass range 
$1.0 \GeVcsq <m(3\pi) <1.5 \GeVcsq$ (Fig.~\ref{fig:nnvar_npm}a).
The $\Lambda$ distribution is consistent with the $J^P=2^+$ expectation
in the $\rho\pi$ decay mode.
The test results for higher mass regions are also consistent 
with the $J^P=2^+$ expectations for the $\rho\pi$ and $f_2 \pi$ modes. 
The $\chi^2$ values obtained are listed in Table~\ref{tab:lambda}.
The good agreement seen in the low $\Lambda$ region indicates 
little contamination by phase space background.
The helicity state is not distinguished by the $\Lambda$ distribution,
but is better determined by the $\cos\theta$ distribution of final 
state pions (to be discussed in Section 6). 

Contributions of known non-tensor resonances,
 $\pi_0(1300) (1^-)$, $\pi_2(1670) (2^-)$, and
$a_4(2040) (4^+)$ are estimated by binned 
maximum likelihood fits to the $\Lambda$ distributions.
The Monte Carlo distributions are generated for a test resonance of 
spin-parity $J^P$ and a tensor resonance (of helicity-2) of mass and 
width listed in Table~\ref{tab:lambda} for the chosen mass range. 
The sum of the distributions, with the bin contents
\begin{equation}
  N^{\MC}_i =  f\cdot N^{\MC}_i(J^P) + (1-f) N^{\MC}_i(2^+) ,
\label{eq:lambdafit}
\end{equation}
is fitted for the fraction $f$. 

The contribution of the $\pi_0(1300)$ is found to be negligible 
in the $\Lambda$ distribution of $m(3\pi) < 1.50$ \GeVcsq{}
for the $\rho\pi$  and $\ppp$ decay modes.
The result $f=0.0\pm 1.1 \pm 1.0$\% is obtained for both decay modes.
The $0^-$ wave is easily distinguishable from the tensor wave and 
therefore a small systematic uncertainty (1.0\%) is estimated 
for the dependence on mass resolution and selection efficiency.
The expectation for the $\pi_0(1300)$ is simulated with the resonance 
parameters from the PDG, $m=1300\pm100$ \MeVcsq{} and width in
the range $\Gamma=200-600$ \MeVcsq.
The number of events estimated for 
$\Ggg(\pi_0(1300))\Br(\pi_0(1300)\to\rho\pi) =1 \eV$ 
is $N^{\MC}=3.24\pm0.32$ for the $\rho\pi$ mode,
and $2.93\pm0.29$ for the $\ppp$ mode.
The errors are estimated from the uncertainties on mass and width.
The upper limit is derived from the $\pi_0(1300)$ 
event fraction found in the fit and the quadratic sum of 
statistical and systematic errors (1.49\%).
It corresponds to 173 of the total 11654 data events.
Applying Poisson statistics, the upper limit at the 90\% confidence 
level (CL) is 191 events, corresponding to two-photon radiative widths 
$\Ggg(\pi_0(1300))\Br(\pi_0(1300)\to\rho\pi) < 65 \eV$ and 
$\Ggg(\pi_0(1300))\Br(\pi_0(1300)\to\ppp) < 72 \eV$,
respectively. 
The $N^{\MC}$ used in the calculation is reduced by one 
standard deviation to account for the uncertainties 
of the $\pi_0(1300)$ mass and width.

Likewise, the contribution of $\pi_2(1670)$ decaying into 
$\rho\pi$ and $f_2\pi$ is determined with the $\Lambda$ distribution 
of 8836 data events satisfying $1.5 \GeVcsq < m(3\pi) <1.85 \GeVcsq$.
The event fraction obtained for the $\pi_2(1670)$ is 
$f=0.0\pm1.2 \pm 2.0$\%.
The expected number of events for 
$\Ggg(\pi_2(1670))\Br(\pi_2(1670)\to\rho\pi,f_2\pi)=1 \eV$ is
$19.3\pm 0.4$ events for $\pi_2(1670)$ with
$m=1672.3\pm3.2$ \MeVcsq{} and $\Gamma=259\pm9$ \MeVcsq.
The upper limit obtained is 
$\Ggg(\pi_2(1670))\Br(\pi_2(1670)\to\rho\pi,f_2\pi) < 12 \eV$ (90\% CL).

The contribution of $a_4(2040)$ is determined for 
helicity 0 and 2 states with the 
$\Lambda$ distribution of 4400 data events with 
$1.85 \GeVcsq < m(3\pi) < 2.1 \GeVcsq$.
The fractions obtained in this case are also negligible, 
with $f=0.0\pm4.5\pm4.0$\% for both helicity states.
The numbers of events expected for 
$\Ggg(a_4(2040))\Br(a_4(2040)\to\rho\pi,f_2\pi) =1$ \eV{}
is $48.2\pm2.2$ and $46.4\pm2.1$ for helicity 0 and 2, respectively.
The upper limit on the helicity 2 wave is found to be 
$\Ggg(a_4(2040))\Br(a_4(2040)\to\rho\pi,f_2\pi) < 6.5 \eV$ (90\% CL).

\section {5. Selection of tensor states }

\begin{figure}[b!]
  \centering\epsfig{file=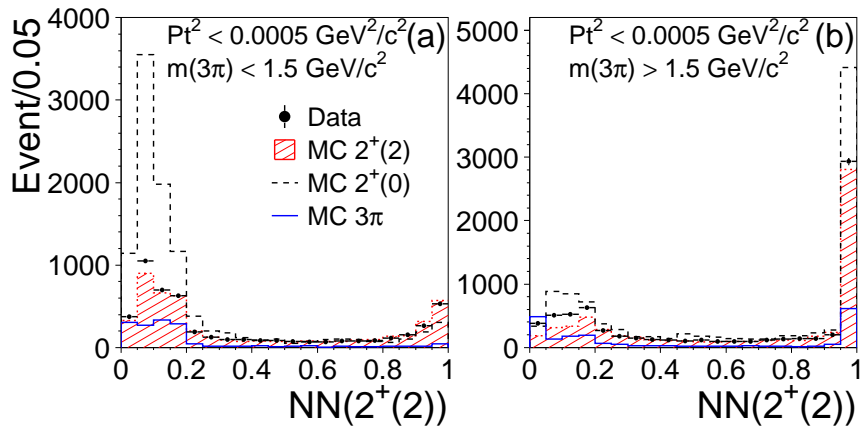,width=.70\linewidth}
  \vspace{-.3cm}
  \caption{ Neural network output for $J^P=2^+$ helicity 2 wave.
     The Monte Carlo samples are normalized to fixed radiative widths
     for comparison of selection efficiencies of different partial 
     waves. $\Ggg\Br=0.7\keV$ is applied (for $a_2(1320)$) in (a),
     and $\Ggg\Br=0.5 \keV$ in (b), respectively.
  \label{fig:nnres2} }
\end{figure}

\subsection {5.1. Neural network method }

A feed-forward neural network is employed to improve the selection 
purity of the chosen partial waves and decay modes.
The neural network contains seven input nodes
and multiple outputs, each corresponding to a partial wave.  
The input variables include the di-pion masses and the $\cos\theta$ 
values of the final state pions and of the vector normal to the 
decay plane.  The three-pion events are best described 
by a tensor partial wave in the $\rho\pi$ and $f_2\pi$ modes.
The neural network is therefore trained to discriminate 
helicity 0, helicity 2, and background (approximated by phase space). 
The product of the outputs of three nodes $O_i$, for 
$J^P(\lambda)=2^+(2)$, $2^+(0)$, and phase space, 
gives the neural network weight
\begin{equation}
   \NN = O_i \cdot (1-O_j) \cdot (1-O_k),
\end{equation}
where $i$ is the wave chosen to be identified against the other two.
Distributions of $\NN$ for helicity 2  are shown in 
Fig.~\ref{fig:nnres2}.
The data sample used has $p_t^2(3\pi) < 0.0005 \GeVsqcsq$
and photon energy $E_\gamma>180$ \MeV{}.
The sample is divided into masses below and above 1.5 \GeVcsq{} 
for comparison of spectra with events dominated by 
the $a_2(1320)$ and those in the higher mass region. 
The $\NN$ spectra are in good agreement with the helicity 2 
(hatched area) predictions.
Note that the detector acceptance is higher for the helicity 0 sample 
(dashed line).
The selection efficiency for the phase space sample is comparable with
that for 
the tensor states; however, the event rate differs by
the spin factor $(2J+1)$ in the cross section. 
With a cut applied at $\NN>0.2$, phase space events at lower mass 
are suppressed by 80\%, leaving about equal numbers of events 
in the helicity 0 and 2 states.
The $a_2(1320)$ selection efficiencies become 0.023\%, 0.053\% and
0.054\% for phase space, tensor helicity 0 and tensor helicity 2 decays,
respectively.

\begin{figure}[ht!]
  \hspace{1.cm}
  \centering\epsfig{file=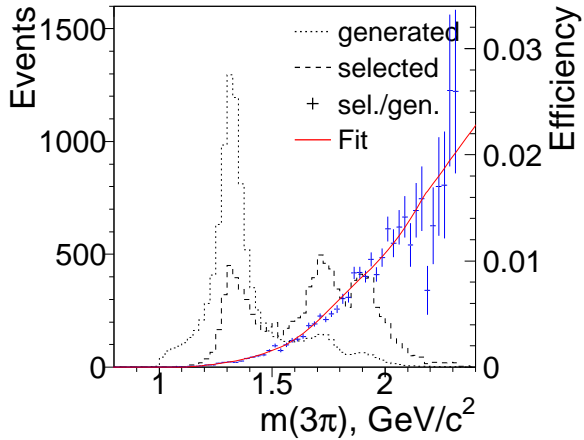,width=.48\linewidth}
  \vspace{-.3cm}
  \caption{Estimated selection efficiency for the
     $J^P=2^+$ helicity 2 partial wave.
    Distributions are shown for Monte Carlo generation
    and fully simulated and selected events.
    The efficiency, shown as points with error bars, is fitted to a 
    polynomial function.
  \label{fig:meff} }
\end{figure}

The event selection cuts were optimized for selection purity.
Because events from three-pion resonances are densely distributed 
at $p_t(3\pi)$ near zero,  
a $p_t^2(3\pi)$ cut is effective in improving selection purity.
A relatively high photon energy threshold suppresses beam related 
background and an $\NN$ threshold suppresses random three-pion 
background.
A data sample enriched in events from tensor states is selected with 
$p_t^2(3\pi)<0.0005 \GeVsqcsq$, $E_\gamma>180 \MeV$, and 
$\NN(2^+(2)) > 0.2$.
The selection efficiency is evaluated with Monte Carlo events.
Figure~\ref{fig:meff} shows the generated and selected mass spectra 
for the $J^P=2^+$ helicity 2 wave in the $\rho\pi$ and $f_2\pi$ modes 
for three resonances at 1320, 1750 and 1950 \MeVcsq{}.
The selection efficiency as a function of $m(3\pi)$ is a smooth curve.
The shape does not change much as the cut values are varied,
which helps in reducing the systematic uncertainty.

\subsection {5.2. Background template }
\label{sec:pt2background}

Background contamination is investigated with rejected events.
Their $p_t^2(3\pi)$ distribution is approximately linear,
which differs very much from two-photon events with a $\rho$-pole
form factor peaking at zero.
The background fraction is estimated from a fit to the $p_t^2(3\pi)$ 
distribution of Monte Carlo two-photon events combined with a
linear background function.

\begin{figure}[h!]
  \centering
  \hspace{-1.0cm}
  \epsfig{file=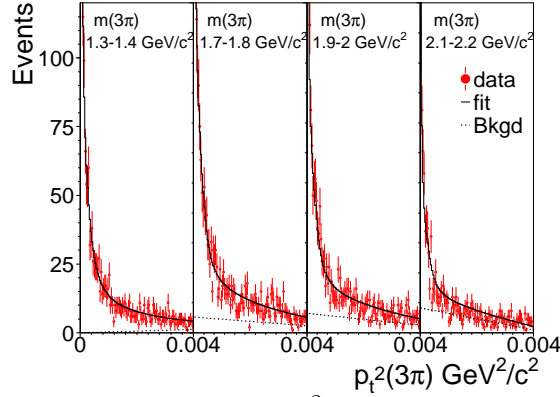,width=.45\linewidth}

  \vspace{-.5cm}
  \caption {Some $p_t^2(3\pi)$ distributions in 100 \MeVcsq{} 
      mass intervals. Each distribution is fitted
      to the Monte Carlo simulation and a linear background.
  \label{fig:mnpt2_pt2} }
\end{figure}

The sample is divided into 100 \MeVcsq{} mass intervals in
$m(3\pi)$; some of the fits to $p_t^2$ distributions are illustrated
in Fig.~\ref{fig:mnpt2_pt2}.
The background fractions are then obtained for a chosen $p_t^2(3\pi)$ 
threshold, and are parameterized as a function of $m(3\pi)$.
The background distribution thus obtained is a smooth curve
corresponding to 15\% of the final sample of 9291 events
(Section 5.1) in the three-pion mass range up to 3 \GeVcsq{}.

\begin{figure}[t!]
  \centering
  \epsfig{file=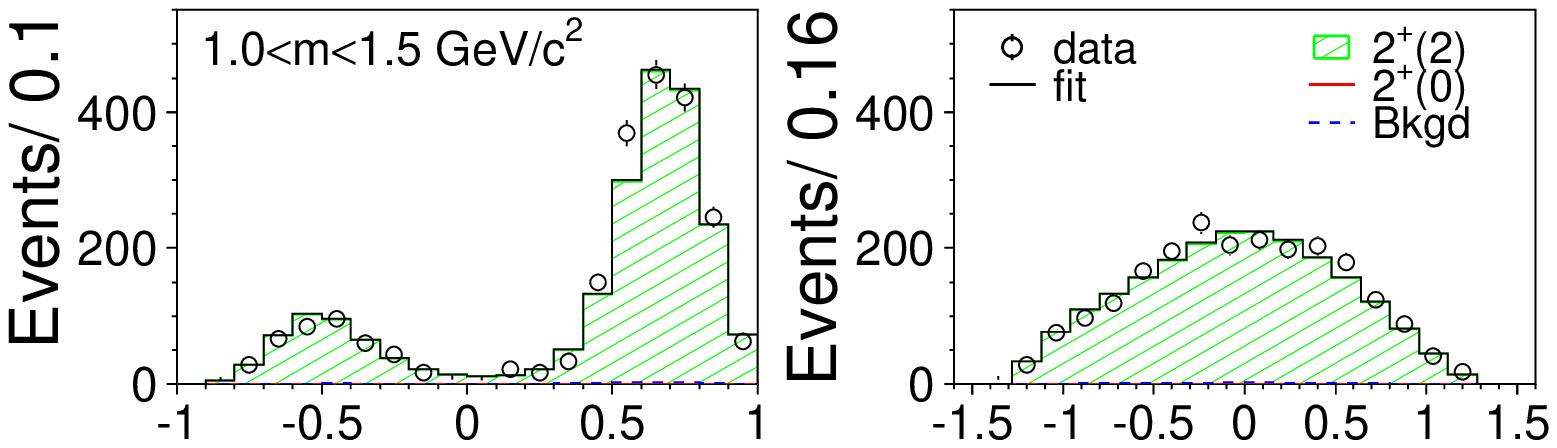,width=.55\linewidth}

  \vspace{-.8cm}
  \epsfig{file=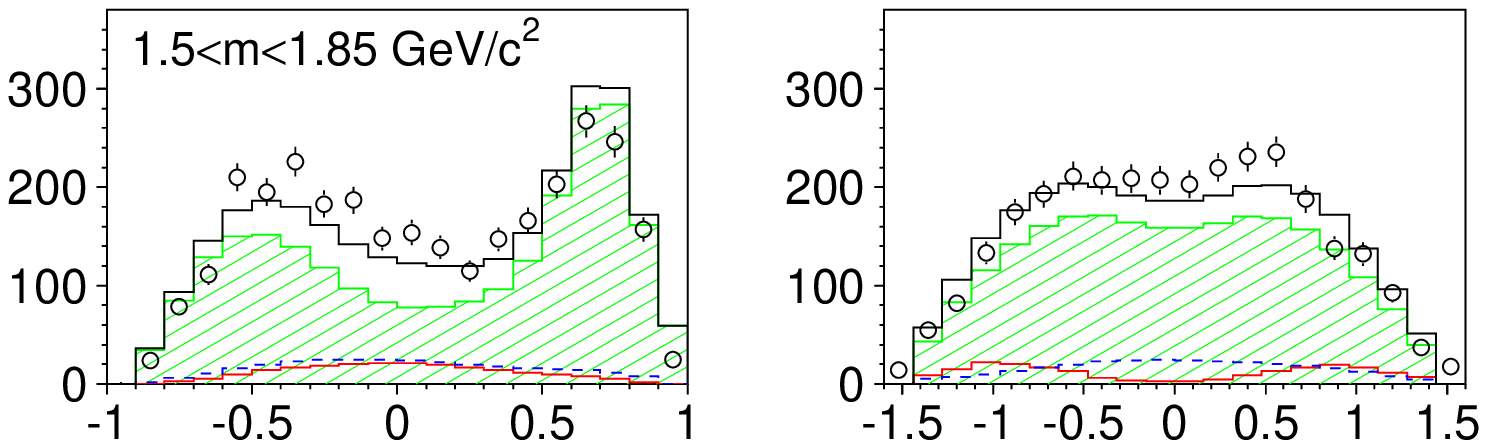,width=.55\linewidth}

  \vspace{-.8cm}
  \epsfig{file=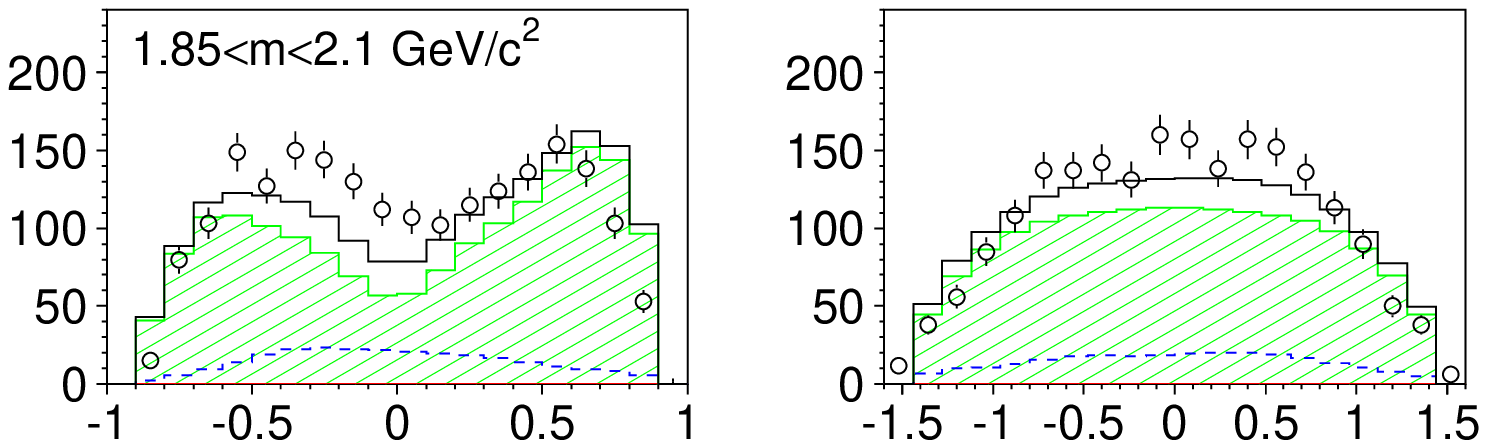,width=.55\linewidth}

  \vspace{-.8cm}
  \epsfig{file=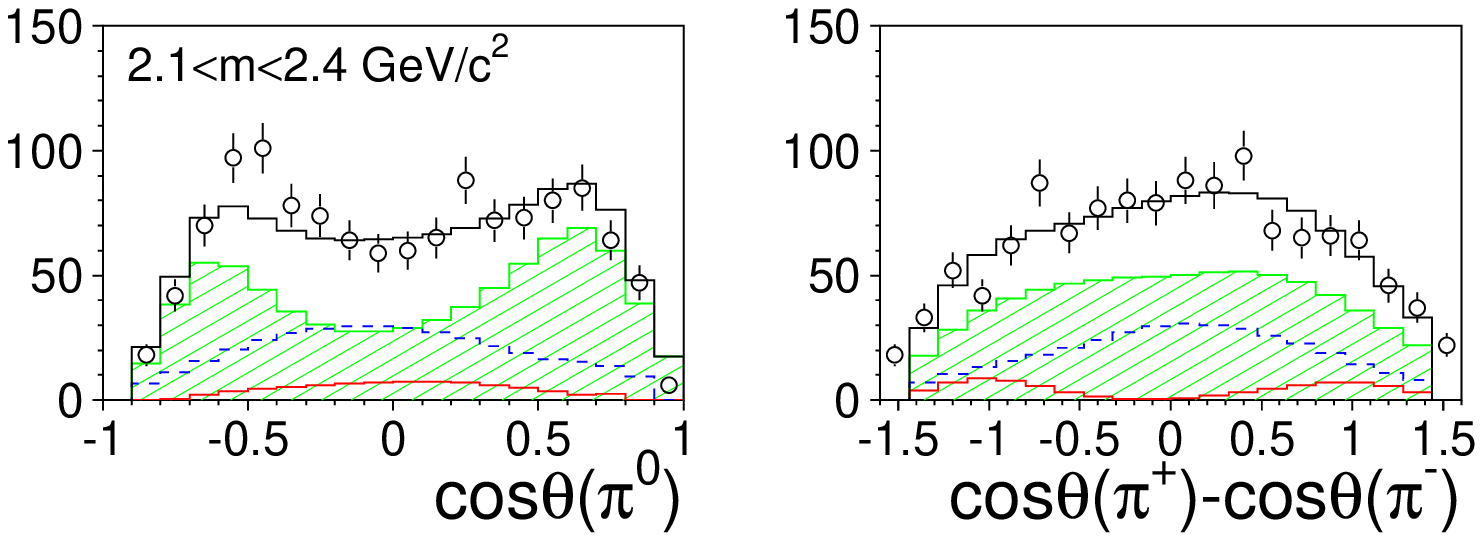,width=.55\linewidth}

  \vspace{-.5cm}
  \caption{ $\cos\theta$ distributions of final state pions.
     Solid lines are the fits of helicity fractions.
  }\label{fig:cosq_loose}
\end{figure}

\section {6. Helicity amplitude }

The helicity fractions of the selected three-pion events
are investigated by examining the polar angle distributions of pions.
The $\cos\theta$ distributions for the of $\pi^0$ and the difference 
in $\cos\theta$ of $\pi^+$ and $\pi^-$ are shown in 
Fig.~\ref{fig:cosq_loose}, in four invariant mass intervals,
corresponding to $a_2(1320)$ and possible higher mass states.
The two helicity states are clearly distinguishable in 
$\cos\theta(\pi^0)$ with helicity 2 (0) events distributed in the 
forward (central) region, and {\it vice versa} for the distributions 
of $\cos\theta(\pi^+) - \cos\theta(\pi^-)$.
The asymmetric shape observed in $\cos\theta(\pi^0)$ distributions
is due to the Belle configuration, which has asymmetric acceptance 
along and opposite to the boost direction.

The helicity fractions are obtained by fits to the $\cos \theta$ 
distributions for the two helicity states and background.
The fits, shown as histograms in 
Fig.~\ref{fig:cosq_loose},
are consistent with the hypothesis of helicity 2 dominance
in all the mass intervals.
The fits with the background fractions fixed to the estimates 
in Section 5.2                              
are listed in Table~\ref{tab:cosq}.
Background is assumed to be randomly distributed in three-pion phase 
space.  Statistically compatible results are 
obtained with background fractions set to zero.
In the low mass region, where $a_2(1320)$ is dominant,
the $\cos\theta$ distributions are consistent with pure helicity 2.
In the higher mass region the helicity state is also predominately
helicity 2.
The Monte Carlo includes multiple tensor states 
in the $\rho\pi$ and $f_2\pi$ channels.
Note the discrepancy in 
$\cos\theta(\pi^0)$ for $1.85 \GeVcsq <m(3\pi)<2.1 \GeVcsq$,
where the Monte Carlo suggests more energetic final state pions
and therefore a more symmetric distribution. 
Systematic uncertainties are estimated by varying $m(3\pi)$ 
intervals, selection cuts and background levels.
The deviation in helicity fractions is evaluated and
the systematic errors estimated are listed in Table~\ref{tab:cosq}.

\begin{table}[t!]
  \begin{center} \begin{tabular}{l|ccc}
  \hline  $m(3\pi)$ mass range
   &  background (fixed) \%      &  Helicity-2 \%   &  $\chi^2$/ndf  \\
  \hline
   1.0 $-$ 1.5  \GeVcsq
   & 0.8  &  $100\pm2 \pm5$   &  40/35   \\
   1.5 $-$ 1.85 \GeVcsq
   & 10   &  $95 \pm2 \pm5$   &  96/37   \\
   1.85 $-$ 2.1 \GeVcsq
   & 13   &  $100\pm2 \pm5 $  &  131/36   \\
   2.1  $-$ 2.4 \GeVcsq
   & 30    & $ 82\pm2 \pm5 $  &  39/37   \\
  \hline
  \end{tabular}
  \caption{ Helicity 2 fractions of the fits to $\cos\theta$ 
     distributions.
     Background fractions estimated in Section 5.2
     are fixed in the fits.
  \label{tab:cosq} }
  \end{center}
\end{table}

\section {7. Mass spectrum of tensor states }

\subsection {7.1. Resonance parameters }
\label{sec:respara}

\begin{figure}[t!]
 \vspace{-.5cm}
 \centering\epsfig{file=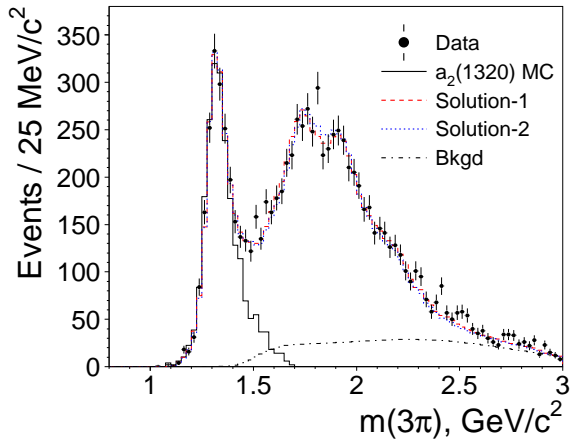,width=.48\linewidth}
  \vspace{-.5cm}
  \caption{Three-pion mass spectrum.
    The fit to $a_2(1320)$ in the mass region below 1.5 \GeVcsq{}
    is shown by the solid line.
    The two solutions of the fit to interference of tensor states are 
    also shown.  The dashed-dotted line is the estimated background.
  }\label{fig:res4_pretty}
\end{figure}

The invariant mass spectrum of three-pion events is shown in 
Fig.~\ref{fig:res4_pretty}.  
The $a_2(1320)$ is dominant in the low mass region where the mass 
spectrum is truncated by trigger thresholds and selection cuts.
The $a_2(1320)$ radiative width is determined by the number of events 
observed in the mass region below 1.5 \GeVcsq.
The Monte Carlo prediction has included full detector simulation
with the $a_2(1320)$ resonance parameters fixed to 
the PDG values ($m=1319 \MeVcsq, \Gamma=105$ \MeVcsq) \cite{PDG}.
The background estimated in Section 5.2 is included.
By scaling the Monte Carlo expectation for $a_2(1320)$ shown by the 
solid line in Fig.~\ref{fig:res4_pretty}, the two photon radiative 
width obtained is
\[  \Ggg(a_2(1320)) =   0.99 \pm0.03 \pm0.11 \keV. \] 
It is in good agreement with the PDG world average.
The systematic uncertainty is estimated to be 11\%.
It is attributed to the trigger efficiency and event selection (7\%)
and the background and contribution from higher mass states (8\%).

The spectrum in the higher mass region shows a structure too broad to 
be due only to the $a_2(1700)$, with width about 200 \MeV{}~\cite{PDG}.
The quick rise at 1.5 \GeVcsq{} and the dip at 
1.8 \GeVcsq{} are indications of overlapping resonances.
The interference of tensor resonances is parameterized in 
Eq.~\ref{eq:xsec} by the coupling amplitudes and phase angles 
relative to the ground state $a_2(1320)$.
The fit for the resonance parameters was conducted with 
the mass spectrum generated by Monte Carlo simulation.
The $a_2(1320)$, serving as the reference ground state, is generated
with the resonance parameters given by the PDG and the resulting 
spectrum is normalized to data in the mass region below 1.5 \GeVcsq{}.

The fit giving satisfactory agreement with the data is achieved
with four resonances, which results in eight sets of solutions.
The interference among the tensor states has the effect of shifting
resonance mass positions and shapes. 
The coupling to the phase angles is periodic with period $2\pi$.
In each half period, one solution is obtained with the amplitudes of 
higher mass states smaller than 1 and closely grouped phase angles.
These solutions are shown by the dashed and dotted lines in 
Fig.~\ref{fig:res4_pretty} for Solution-1 and 2, respectively,
with $\chi^2/ \ndf = 45/50$ and 46/50 in the mass range up to 
$2.4 \GeVcsq$.
The parameters obtained are listed in Table~\ref{tab:m4ptst_b}.
The solutions with interference amplitudes larger than 1
(Solution$(\alpha_i>1$), Table~\ref{tab:m4ptst_b})
correspond to destructive interference with radially excited states 
having larger cross sections than the ground state ($a_2(1320)$), 
which is unlikely for light-quark mesons.

\begin{table}[b!]                         
  \begin{center} \begin{tabular}{c|cccc|c} 
  \hline  
  Solution-1     & mass \MeVcsq          &  width \MeVcsq  
                 & amplitude ($\alpha$'s) 
                 & phase ($\phi$'s) deg. & $\chi^2$/ndf \\
  \hline
  $a_2(1700)$    & $1769\pm10    \pm8$     & $270\pm7  \pm10$  
                 & $0.37 \pm0.02 \pm0.03 $ & $154\pm6  \pm12$ & 45/50 \\
  $BW(1950)$     & $1948\pm4     \pm8$     & $291\pm6  \pm10$
                 & $0.61 \pm0.02 \pm0.06 $ & $143\pm5  \pm12$  \\
  $BW(2140)$     & $2146\pm12    \pm 8$    & $358\pm26 \pm10$
                 & $0.40 \pm0.03 \pm0.04 $ & $139\pm4  \pm12$ \\
  \hline
  Solution-2     & mass \MeVcsq         &  width \MeVcsq  
                 & amplitude ($\alpha$'s) 
                 & phase ($\phi$'s) deg.\\
  \hline
  $a_2(1700)$    & $1758\pm13    \pm8$     & $269\pm10 \pm10$
                 & $0.37 \pm0.03 \pm0.03 $ & $221\pm7  \pm12$ & 46/50 \\
  $BW(1950)$     & $1949\pm6     \pm8$     & $324\pm14 \pm10$
                 & $0.71 \pm0.03 \pm0.06 $ & $220\pm6  \pm12$  \\
  $BW(2140)$     & $2161\pm17    \pm 8$    & $342\pm22 \pm10$
                 & $0.44 \pm0.03 \pm0.04 $ & $221\pm6  \pm12$ \\
  \hline
  \multicolumn{5}{c}{ } \\
  \hline
  Solution($\alpha_i>1$) 
                 & mass \MeVcsq         &  width \MeVcsq
                 & amplitude ($\alpha$'s) 
                 & phase ($\phi$'s) deg. &  $\chi^2$/ndf  \\
  \hline  
  $a_2(1700)$    & $1768\pm5     \pm8$     & $273\pm 8  \pm10$
                 & $0.50 \pm0.02 \pm0.05 $ & $126\pm 2  \pm 8$ & 48/50\\
  $BW(1950)$     & $1949\pm3     \pm8$     & $280\pm 4  \pm10$
                 & $1.54 \pm0.02 \pm0.14 $ & $ 99\pm 1  \pm 8$  \\
  $BW(2140)$     & $2138\pm4     \pm 8$    & $382\pm 5  \pm10$
                 & $2.03 \pm0.03 \pm0.18 $ & $236\pm 1  \pm 8$ \\
  \hline  
  $a_2(1700)$    & $1757\pm 6    \pm8$     & $279\pm12  \pm10$
                 & $0.54 \pm0.02 \pm0.05 $ & $207\pm 3  \pm 8$ & 49/50\\
  $BW(1950)$     & $1954\pm 3    \pm8$     & $289\pm 5  \pm10$
                 & $1.64 \pm0.02 \pm0.15 $ & $163\pm 1  \pm 8$  \\
  $BW(2140)$     & $2136\pm 4    \pm 8$    & $374\pm 4  \pm10$
                 & $2.23 \pm0.03 \pm0.20 $ & $313\pm 1  \pm 8$ \\
  \hline  
  $a_2(1700)$    & $1750\pm 4    \pm8$     & $271\pm 3  \pm10$
                 & $1.45 \pm0.02 \pm0.13 $ & $102\pm 1  \pm 8$ & 49/50\\
  $BW(1950)$     & $1960\pm 4    \pm8$     & $267\pm 5  \pm10$
                 & $1.02 \pm0.02 \pm0.09 $ & $237\pm 1  \pm 8$  \\
  $BW(2140)$     & $2138\pm 9    \pm 8$    & $375\pm14  \pm10$
                 & $0.64 \pm0.02 \pm0.06 $ & $212\pm 2  \pm 8$ \\
  \hline  
  $a_2(1700)$    & $1759\pm 4    \pm8$     & $269\pm 4  \pm10$
                 & $1.70 \pm0.02 \pm0.15 $ & $161\pm 1  \pm 8$ & 51/50\\
  $BW(1950)$     & $1963\pm 4    \pm8$     & $263\pm 5  \pm10$
                 & $1.31 \pm0.02 \pm0.12 $ & $305\pm 2  \pm 8$  \\
  $BW(2140)$     & $2136\pm 9    \pm 8$    & $360\pm12  \pm10$
                 & $0.76 \pm0.03 \pm0.07 $ & $290\pm 3  \pm 8$ \\
  \hline  
  $a_2(1700)$    & $1745\pm 2    \pm8$     & $268\pm 2  \pm10$
                 & $1.88 \pm0.02 \pm0.17 $ & $ 81\pm 1  \pm 8$ & 49/50\\
  $BW(1950)$     & $1947\pm 2    \pm8$     & $275\pm 2  \pm10$
                 & $2.78 \pm0.02 \pm0.25 $ & $188\pm 1  \pm 8$  \\
  $BW(2140)$     & $2143\pm 2    \pm 8$    & $372\pm 4  \pm10$
                 & $2.29 \pm0.02 \pm0.21 $ & $305\pm 1  \pm 8$ \\
  \hline  
  $a_2(1700)$    & $1742\pm 2    \pm8$     & $263\pm 2  \pm10$
                 & $1.94\pm 0.02 \pm0.17 $ & $143\pm 1  \pm 8$ & 51/50\\
  $BW(1950)$     & $1944\pm 2    \pm8$     & $263\pm 2  \pm10$
                 & $2.81\pm 0.02 \pm0.25 $ & $257\pm 1  \pm 8$  \\
  $BW(2140)$     & $2148\pm 2    \pm 8$    & $378\pm 5  \pm10$
                 & $2.42\pm 0.03 \pm0.22 $ & $ 20\pm 1  \pm 8$ \\
  \hline
  \end{tabular}
  \caption{Resonance parameters of the fit to the invariant mass
      spectrum for tensor states with interference expressed in 
      Eq.~\ref{eq:xsec}.
  \label{tab:m4ptst_b} }
  \end{center}
  \vspace{-.5cm}
\end{table}

Correlations between the fitted phase angles and resonance parameters
are significant.
The resonance mass positions are shifted to values higher than in the 
fit without interference.  
The largest correlation coefficient
is between $\phi_1$ and $m(1700)$ for $a_2(1700)$.
The values obtained are -0.72 (-0.88) in Solution-1 (2), respectively.
The width of the resonance is positively correlated with the 
interference amplitude.
The relative amplitude $\alpha_2$ has a large correlation coefficient
with the width of the resonance at 1950 \MeVcsq.
The values are 0.74 (0.75) in Solution-1 (2), respectively.
The interference of overlapping resonances leads to  strong 
correlations and to shifts in two-photon radiative widths.
The radiative width of $a_2(1700)$ in three-pion mode is found using
$\Ggg(1700)\Br(3\pi) =\alpha_1^2\cdot\Ggg(1320)\Br(\rho\pi)$
to be $0.10 \pm 0.02\pm0.02$ keV.    

Systematic uncertainties are evaluated including the effects of 
interference on selection efficiency and mass resolution.
The mass resolution is smeared in the reconstruction of $\pi^0$  
and charged pion tracks.
The smearing effect on mass is estimated by Monte Carlo and is found 
to be approximately equivalent to convolution with a Gaussian of 
width less than 20 \MeVcsq{}.
The Monte Carlo spectrum is slightly shifted with respect to data and 
the magnitude of the shift is examined as a function of the trigger 
and event selection thresholds.
The precision of the resonance mass measurement 
is estimated to be better than 8 \MeVcsq.

The uncertainty in the resonance width is due to the smearing of the 
invariant mass measurement.
It is estimated to be 10 \MeVcsq{} by comparing the $p_t(3\pi)$ of 
data and Monte Carlo and by comparing the simulated and measured widths
of the $a_2(1320)$.
The systematic error from interference amplitudes is 9\%.
It is estimated using the uncertainties of selection efficiencies 
and backgrounds.
The interference phase angles are strongly correlated with the 
resonances masses.  
The systematic error on masses corresponds to an error of 8 degrees
for the phase angles in Solution-1 and 2.
In addition, the phase angles are affected by the decay modes,
and the errors are evaluated in the fits to be 8 degrees.
The combined error is estimated to be 12 degrees.
The fits having large amplitudes $(\alpha_i>1)$ are more sensitive to
phase angle (smaller statistical error) and are less dependent on
the errors in the masses.
The systematic uncertainty of the phase angles is thus smaller.

\begin{figure}[b!]
  \vspace{-.2cm}
  \centering
  \epsfig{file=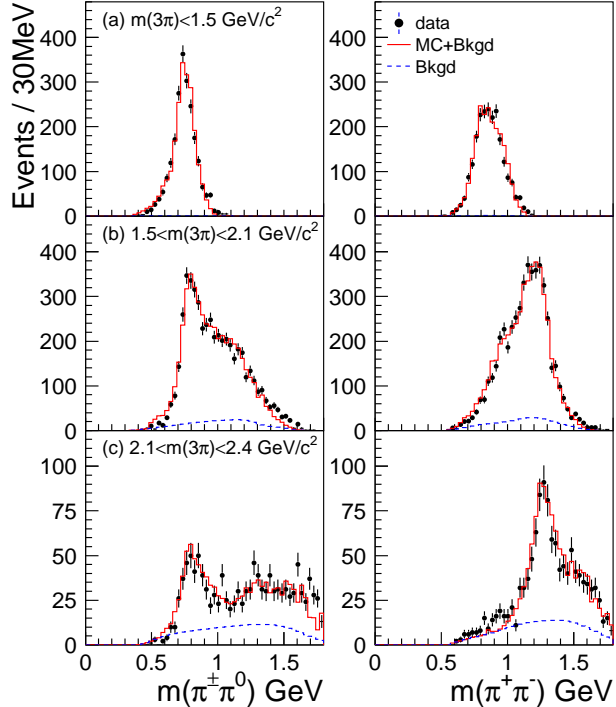,width=.52\linewidth}
  \vspace{-.5cm}
  \caption{ Di-pion mass spectra and the fits
   to $\rho\pi$ and $f_2\pi$ modes.
   The interference parameters are listed in Table~\ref{tab:decay_int}.
  }\label{fig:a2aphi_t} 
\end{figure}

\subsection {7.2. Intermediate decay channels }
\label{sec:decay}

The presence of $\rho\pi$ and $f_2\pi$ intermediate states
is demonstrated by the di-pion invariant mass spectra shown in 
Fig.~\ref{fig:a2aphi_t}.
The $a_2(1320)$ decays into $\rho\pi$ only.
The $m(\pi^\pm\pi^0)$ spectrum for $m(3\pi)< 1.5 \GeVcsq$
(Fig.~\ref{fig:a2aphi_t}a) is consistent with simulated 
$\rho\pi$ decay and the $m(\pi^+\pi^-)$ spectrum from phase space.
The $f_2$ is present in the $m(\pi^+\pi^-)$ distributions of
$m(3\pi) >1.5$ \GeVcsq.
The interference between $\rho\pi$ and $f_2\pi$ modes is formulated 
with an amplitude and a phase angle in Eq.~\ref{eq:di}.
The tensor resonance parameters have little effect 
on the di-pion invariant mass spectra and the final state pion 
distributions.
Therefore the decay parameters are determined independently.

The interference term has the effect of pulling di-pion mass peaks for 
$\rho$ and $f_2$, near their nominal mass positions for 
$\psi=180^\circ$, to give destructive interference and flatter 
spectra as $\psi$ approaches zero.
The interference parameters are determined by a least-squares 
fit to the two di-pion mass spectra.
The Monte Carlo expectation includes the resonances in Section 7.2.
To account for the correlation between resonances,
the fit was performed for the combined sample of $a_2(1700)$ 
and the resonance at 1950 \MeVcsq{} with 
$1.5 \GeVcsq < m(3\pi) < 2.1 \GeVcsq$.
The background included is a phase space distribution scaled to
the fraction estimated in Section 5.2.  
The fit obtained with $\chi^2/ \ndf = 92/51 $ 
is illustrated in Fig.~\ref{fig:a2aphi_t}b.
The parameters obtained are listed in Table~\ref{tab:decay_int}.
The interference amplitudes and phase angles are similar for
the two resonances.
The fit was also conducted for the spectra in narrower
$m(3\pi)$ intervals corresponding to the dominance of $a_2(1700)$
and the resonance at 1950 \MeVcsq{}.  The Monte Carlo was
simulated with the event fractions of resonances varied.
The results are statistically compatible
and are used to estimate systematic uncertainty.

The event fraction attributed to the resonance at 2140 \MeVcsq{}
is only about a quarter of the total events
in the mass range  $2.1 \GeVcsq < m(3\pi) < 2.4 \GeVcsq$.
The interference parameters are again determined by fitting the di-pion 
mass spectra.  The $\chi^2/ \ndf = 63/59$ is obtained in the fit 
(Fig.~\ref{fig:a2aphi_t}c).
The parameters obtained are also listed in Table~\ref{tab:decay_int}.  

The large overlap of events in adjacent tensor states
leads to large systematic uncertainty.
To estimate the resulting uncertainties, fits to the di-pion mass 
spectra of events selected in wider mass intervals of $m(3\pi)$ were
carried out.
For the resonance at 2140 \MeVcsq{},
the error is up to 30\% for the amplitude and 20\% for the angle.
The estimated errors  are included in Table~\ref{tab:decay_int}.  

\begin{table}[t!]
  \begin{center} \begin{tabular}{c|cc}
  \hline  
                 & amplitude ($\xi$'s) &  phase ($\psi$'s) deg.  \\
  \hline
  $a_2(1700)$    & $0.92\pm0.10\pm0.08$ & $151\pm4 \pm12$  \\
  $BW(1950)$     & $0.91\pm0.10\pm0.12$ & $149\pm4 \pm20$  \\
  $BW(2140)$     & $1.0\pm0.20\pm0.30$ & $145\pm10 \pm30$  \\
  \hline
  \end{tabular}
  \caption{Interference parameters for $\rho\pi$ and
     $f_2\pi$ decay modes of the fit to the di-pion mass spectra.
  \label{tab:decay_int} }
  \end{center}
  \vspace{-.5cm}
\end{table}

\section {8. Conclusion} 

In the region of low three pion mass the reaction $\gamma\gamma\to \ppp$
is dominated by the $a_2(1320)$ in the helicity 2 state.
Events in higher mass regions are also found to be 
dominated by the spin-parity $J^P=2^+$ helicity 2 partial wave.
The contribution of $\pi_0(1300)$ is negligible. 
The upper limit is found to be 
$\Ggg(\pi_0(1300))\Br(\pi_0(1300)\to\ppp) < 72 \eV$ 
(90\% C.L).
The contributions of the $\pi_2(1670)$ and $a_4(2040)$ resonances
are also negligible. 
The upper limits are determined to be 
$\Ggg(\pi_2(1670))\Br(\pi_2(1670)\to\rho\pi,f_2\pi) < 12 \eV$   and
$\Ggg(a_4(2040))\Br(a_4(2040)\to\rho\pi,f_2\pi) < 6.5 \eV$ (90\% CL),
respectively.

The observed radiative width of the $a_2(1320)$ is consistent with 
the PDG world average.
The mass spectrum in the region above the $a_2(1320)$
consists of a broad distribution attributed to the $a_2(1700)$ 
and is best interpreted with two more radially excited tensor states 
at 1950 \MeVcsq{} and 2140 \MeVcsq{} decaying to $\rho\pi$ and $f_2\pi$.
The mass spectrum is well represented by interfering tensor states
with coupling amplitudes and phase angles relative to $a_2(1320)$.
The amplitude obtained for $a_2(1700)$ is 0.37, corresponding
to the radiative width of 
$\Ggg(a_2(1700))\Br(a_2(1700)\to\rho\pi,f_2\pi)
= 0.10 \pm0.02 \pm 0.02 \keV$.

The higher mass tensor states observed here may be compared to the
theoretical predictions of the relativistic quark model 
\cite{Cahn,Ackleh,Munz}.
The radiative widths obtained in this study are compatible with
the relativistic quark model calculations.


{}


\begin{thebibliography}{99}

  \bibitem{Jade83}
     JADE Collaboration, J.E. Olsson, in  Proc. V Int. Conf.  
     on Two-Photon Physics, Aachen 1983 ed. Ch. Berger 
     (Springer, Berlin 1983).

  \bibitem{CELLO83}
     H.J. Behrend {\it et al.}
     (CELLO Collaboration),
     Phys. Lett. 114B (1982) 378;
     Phys. Lett. 125B (1983) 518.

  \bibitem{Pluto84}
     Ch. Berger {\it et al.}
     (PLUTO Collaboration),
     Phys. Lett. 149B (1984) 427.

  \bibitem{Toss86}
     M. Althoff {\it et al.}
     (TASSO Collaboration),
     Z. Phys. C 31 (1986) 537.

  \bibitem{Mark290}
     F. Butler {\it et al.}
     (MARK2 Collaboration),
     Phys. Rev. D 42 (1990) 1368.

  \bibitem{MD1}
     S.E. Baru {\it et al.}
     (MD1 Collaboration),
     Z. Phys. C 48 (1990) 581.

  \bibitem{CELLOpi2}
     H.J. Behrend {\it et al.}
     (CELLO Collaboration),
     Z. Phys. C 46 (1990) 583.

  \bibitem{CBALpi2}
     D. Antreasyan {\it et al.}
     (Crystal Ball Collaboration),
     Z. Phys. C 48 (1990) 561.

  \bibitem{ARGUS}
     H. Albrecht {\it et al.}
     (ARGUS Collaboration),
     Z. Phys. C 74 (1997) 469.

  \bibitem{L3}
     M. Acciarri {\it et al.}
     (L3 Collaboration), 
     Phys. Lett. B 413 (1997) 147.

  \bibitem{Cahn}
     J.D. Anderson, M.H. Austern and R.N. Cahn,
     Phys. Rev. D 43 (1991) 2094.

  \bibitem{Ackleh}
     E.S. Ackleh {\it et al.},
     Phys. Rev. D 45 (1992) 232.

  \bibitem{Munz}
     C.R. M\"{u}nz,
     Nucl. Phys. A 609 (1996) 364.

  \bibitem{Richen}
     R. Richen {\it et al.},
     Eur. Phys. J. A9 (2000) 211.

  \bibitem{Anisovich}
     A.V. Anisovich {\it et al.},
     Phys. Atom. Nucl. 66 (2003) 914.


  \bibitem{BelleKK}
     K. Abe {\it et al.}
     (Belle Collaboration),
     Eur. Phys. J. C 32 (2004) 323.

  \bibitem{PDG}
     Particle Data Group, Phys. Lett. B 592, (2004) 1.

  \bibitem{Antipov}
     Yu.M. Antipov {\it et al.},
     Nucl. Phys. B 63 (1973) 167.

  \bibitem{Deutschmann}
     M. Deutschmann {\it et al.},
     Nucl. Phys. B 114 (1976) 237.

  \bibitem{ACCMOR}
     C. Daum {\it et al.}
     (ACCMOR Collaboration),
     Phys. Lett. 89B (1980) 285; Nucl. Phys. B 182 (1981) 269.

  \bibitem{Aston}
     D. Aston {\it et al.},
     Nucl. Phys. B 189 (1981) 15.

  \bibitem{Condo}
     G.T. Condo {\it et al.} 
     (SLAC Hybrid Facility Photon Collaboration),
     Phys. Rev. D 43 (1991) 2787.

  \bibitem{CBAR02}
     C. Amsler {\it et al.}
     (Crystal Barrel Collaboration), 
     Eur. Phys. J. C23 (2002) 29.
  \bibitem{CBAR99}
     A. Abele {\it et al.},
     (Crystal Barrel Collaboration), 
     Eur. Phys. J. C8 (1999) 67.

  \bibitem{CBAR_AVA}
     A.V. Anisovich {\it et al.}, 
     Phys. Lett. B452 (1999) 187.


  \bibitem{KEKB}
     S.Kurokawa and E.Kikutani, Nucl. Instr. Meth. A 499 (2003) 1,
     and other papers included in this volume.


  \bibitem{Yang}
     C.N. Yang, Phys. Rev. 77 (1950) 242.

  \bibitem{Li}
     Z.P. Li, F.E. Close, T. Barnes, Phys. Rev. D43 (1991) 2161.

  \bibitem{Grassberger}
     see e.g. P. Grassberger and R. Kogerler,
     Nucl. Phys. B 106 (1976) 451.

  \bibitem{Poppe}
     M. Poppe, Int. J. Mod. Phys. 1 (1986) 545.

  \bibitem{Budnev}
     V.M. Budnev {\it et al.},
     Phys. Rep. 15 (1975) 181.


  \bibitem{DIAG36}
     F.A. Berends, P.H. Daverveldt and R. Kleiss,
     Nucl. Phys.  B253 (1985) 421; 
     Comp. Phys. Comm. 40 (1986) 271, 285, and 309. 

  \bibitem{GEANT}
     R.~Brun {\it et al.}, GEANT 3.21 CERN-DD/EE/84-1, 1984.

  \bibitem{belle}
     A. Abashian {\it et al.} (Belle Collaboration),
     Nucl. Inst. and Meth. A 479 (2002) 117.

  \bibitem{ElectronID}
     K. Hanagaki {\it et al.}, Nucl. Instr. Meth. A 485 (2002) 490.



\end{thebibliography}
\end{document}